\documentstyle[11pt,newpasp,twoside,epsf]{article}
\def\edcomment#1{\iffalse\marginpar{\raggedright\sl#1\/}\else\relax\fi}

\reversemarginpar

\begin{document}
\title{Star Formation Modes in Low-Mass Disk Galaxies}
 \author{John S. Gallagher, III}
\affil{Department of 
Astronomy, University of Wisconsin, Madison, WI, USA}
\author{Lynn D. 
Matthews}
\affil{National Radio Astronomy Observatory, Charlottesville, 
VA, USA}

\begin{abstract}
Low-mass disk galaxies with well-organized structures are relatively 
common in low density regions of the nearby Universe. They display 
a wide range in levels of star formation activity, extending from 
sluggishly evolving `superthin' disk systems to nearby starbursts. 
Investigations of this class of galaxy therefore provides opportunities 
to test and define models of galactic star formation processes. In this 
paper we briefly explore characteristics of examples of quiescent and 
starbursting low-mass disk galaxies.
\end{abstract}

\section{Introduction}

Low-mass spiral galaxies exist within the luminosity range
0.1$L^*$-0.01$L^*$, between the standard giant spirals and the typical
dwarf irregulars.  These objects have morphological types of Sc-Sdm,
are reasonably organized, sometimes have thin stellar disks, display
modest spiral structure, and often contain large amounts of HI gas
(e.g., Matthews \& Gallagher 1997).  Under normal conditions their
star formation rates tend to be low, but their substantial gas
contents make them excellent hosts for starbursts.

In this paper we first consider quiescent star formation in low-mass
spiral galaxies as observed in nearby, dynamically cool `superthin' disk
galaxies.  We then turn to properties of high surface brightness, blue
starbursts, which frequently occur in small disk systems, and may be an
important contributor to the large populations of faint blue galaxies
found at moderate redshifts.

\section{Star Formation in Slowly Evolving Galactic Disks}

\subsection{Basic Characteristics}

The low mass Sd spirals include excellent examples of pure disk
galaxies; galaxies which show no evidence for a stellar spheroid. 
Observationally we most readily identify such galaxies
when they are seen near edge-on, allowing us to judge the thickness of
the stellar disk.  Some low-mass spirals have thin stellar disks--i.e,
they are superthin galaxies with little or no stellar halo, and large
ratios of radial to vertical disk scale lengths (Gallagher \& Hudson
1976, Goad \& Roberts 1981, Matthews et al. 1999, Dalcanton \& Bernstein 
2000, Matthews 2000).  These objects are also rich in HI gas, and may have low
metal abundances (Bergvall \& R\"onnback 1995, Giovanelli et al. 1997,
Matthews \& van Driel 2000). Figure 1 compares WIYN Telescope R-band
and H$\alpha$ images of UGC~7321, an outstanding example of a
superthin disk.

\begin{figure}
\begin{center}
\leavevmode
\epsfxsize=3.5in
\epsffile{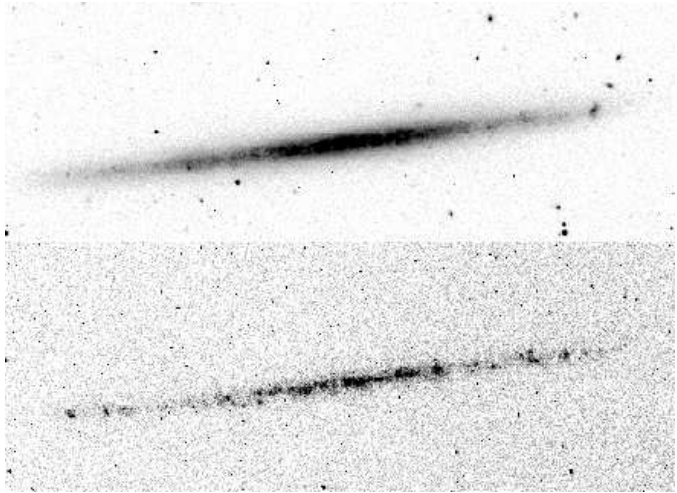}
\end{center}
\vspace{-0.5cm}
\caption{Mosaic showing WIYN Telescope 
R-band and narrow band H$\alpha +$[NII] emission 
line images of the superthin galaxy UGC 7321. }
\end{figure}

This class of galaxy is of special interest since they
would appear to have been assembled from objects containing only gas,
and therefore experienced only minimal star formation before their  
their disks were fully formed. Such galaxies are difficult to make in standard
cold dark matter (CDM) scenarios (Gnedin et al. 2000, 
Wyse 2000). The presence of a thin stellar disk
further indicates that little dynamical heating occurred over
most of the lifetime of the disk, 
while the presence of large gas
content demonstrates that in a global sense star formation has been
inefficient (Matthews 2000).  Low-mass Sd spirals are thus 
the least evolved examples of well-organized
disk galaxies in the nearby universe.

\subsection{Star Formation Patterns in UGC~7321}

Figure 1 shows the overall pattern of star formation in UGC~7321 as
traced by its HII regions. These extend over almost the entire
optically visible disk, but are most frequent and brightest in the
central regions of the galaxy.  Further insight into products of recent
star formation and the state of the ISM come from high angular
resolution {\it HST} WFPC2 images. As seen in Figure 2, the central region of
UGC~7321 looks like a normal galactic disk. 
While there is no continuous dust lane, dusty dark clouds
are obvious and suggestive of the presence of molecular gas, a view
that is supported by the detection of molecular CO 1-0 emission by
Matthews \& Gao (2001).  The {\it HST} images also reveal distinct clusters
and associations of luminous stars in the inner parts of this galaxy.

The outer parts of the UGC~7321 exhibit somewhat different 
characteristics (Figure 2). Dark nebulae are
larger and have lower contrast, while the distribution of luminous
stars becomes more diffuse. Possibly two modes of star formation
exist within the same galaxy. At radii of $<$2~kpc, star formation
appears to occur in a normal cycle of well-defined clouds embedded
within a stellar disk. In this region we could be seeing
self-regulated star formation, where energy supplied by massive stars
limits the overall density of the gaseous disk. Further out, where the
scale height of the gaseous component becomes comparable to that of the
stars, we are probably observing an internally supported gas disk
(e.g., by turbulence; Elmegreen \& Parravano 1994), where star
formation could be playing less of a role in the vertical support of
the ISM. The slow evolution of the outer parts of this galaxy then is a
natural result of inefficient star formation in a diffuse gaseous disk
(see Gallagher et al. 2001 for details).

\begin{figure}
\begin{center} 
\leavevmode
\epsfxsize=3.3in
\epsffile{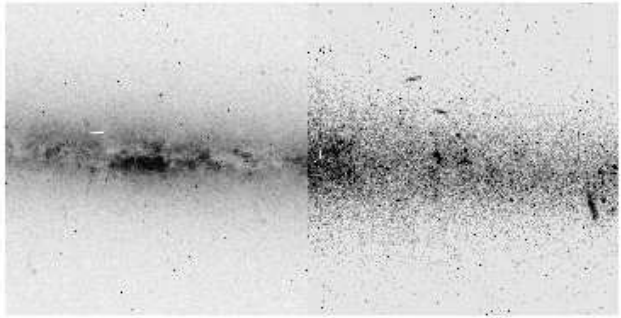}
\end{center}
\caption{{\it HST} WFPC2 images of the center and outer disk of UGC~7321. 
The central regions of this galaxy contain the normal dark clouds and 
clustered star formation, while dark nebulae larger and more diffuse and 
young stars more dispersed in the outer parts of the disk.}
\end{figure}

Gallagher et al. (2001) also estimate a star formation rate from the
number of dark clouds to be $\psi =$10$^{-1.5 \pm 0.5}~M_{\sun} {\rm
yr}^{-1}$. For a stellar mass of 2 $\times$10$^9~M_{\sun}$ the Roberts
time to form the existing mass of stars is roughly a Hubble time.
UGC~7321 is most likely an example of a slowly evolving galaxy rather
than a young galaxy, a model that also is supported by the complex
vertical structure of its stellar disk (Matthews 2000).

\section{Interactions and Starbursts in Small Disks}

\subsection{The Basics}

In the transition region between giant spirals and 
star-forming dwarfs we find galaxies with a variety of structures,
including the Magellanic irregulars with their thick stellar disks and
frequently asymmetric structures, as well as dwarf S0 systems where
star disk star formation appears to be winding down.  Yet in other
ways, such as peak rotation velocities, these galaxies appear to be
similar types of objects. A natural suggestion then is that
interactions join with initial conditions in determining the structures
of the current populations of intermediate luminosity disk 
(and possibly dwarf E; Mayer et al. 2001) galaxies.

To test this hypothesis we consider observational signatures of
interactions and of any possibly associated starbursts:

\noindent {\bf $\diamondsuit$~Interactions} between galaxies directly
perturb the outer regions of the systems which collide, and may yield
wide-spread changes in cases where mergers occur. After-effects may
include an increase in disk vertical scale height, shell or ripple
structures, or tidal tails (Schweizer \& Seitzer 1988, Howard et al. 1993, 
Vel\'azquez \& White 1999). Even a distant collision can suffice to
drive gas inwards thereby triggering a starburst 
(e.g., Mihos \& Hernquist 1994, Barton et al. 2000).

\noindent {\bf $\heartsuit$~Starbursts} occur when there is a major
(factor of $\sim$10) and unsustainable increase in a galaxy's star
formation rate. Ongoing starbursts are obvious from the signatures of
their huge populations of high mass stars, including strong emission
lines, large scale gas outflows, and high power outputs per unit
baryonic mass. Products of starbursts include compact super star
clusters which may be long-lived.

\subsection{Starbursts in Small Disk Galaxies}

In the nearby Universe, 
small galaxies host high surface brightness, compact
starbursts. That these starbursts reside in disk galaxies can be
established from the kinematics and presence of spirals arms and other
kinematic features associated with dynamically cool stellar populations
(e.g., NGC 3310: Mulder et al. 1995; NGC 7673: Homeier \& Gallagher 1999). 
This class
of starburst also includes the nearby archetype M82, where due to its
highly inclined orientation the disky structure of the system is
obvious while much of the starburst is hidden from optical view.

Preliminary results from a small survey of nearby compact starburst
galaxies include evidence for starburst triggering through interactions;
e.g., structural or kinematic features that suggest a minor merger or
recent interaction or the presence of obviously disturbed companions.
In these systems it often appears that a relatively minor perturbation
has led to large scale starbursts. Furthermore, the starbursts
sometimes persist well after the interaction is past; e.g., in
NGC~7673.

A complete picture of the conditions leading to compact starbursts is
not yet in hand. However, it has been known for some time that these
systems frequently contain large amounts of HI (Smoker et al. 2000);
we suspect their gas-rich nature is a key factor in their
susceptibility to starbursts.  Our model is that compact starbursts
take place in gassy disks that were on the margins of stability before
they were perturbed (e.g., galaxies like UGC~7321). The external
perturbation then drives gas inwards, either directly or via the
production of bars or spiral arms, thereby fueling the starburst 
(e.g., Noguchi 1988).

The presence of large amounts of gas also provides a basis for
understanding the large scale clumpy patterns of star formation which
are frequently seen in this class of starburst.  As emphasized by
Elmegreen \& Efremov (1997) and Noguchi (1999), 
a disturbed, gas-rich disk will be
unstable and tend to break up into large bound clumps, which we may see as
huge regions of star formation. Our high angular
resolution observations with HST of NGC~7673, for example, 
reveal that the clumps often are
associations of dense star clusters embedded in backgrounds consisting
of luminous massive stars (see also de Grijs et al. 2001).  
Evidently in these circumstances the star
forming hierarchy has gone one level beyond that found in OB
associations with dense star clusters paralleling the role of the most
massive stars under more normal circumstances.

\subsection{The Initial Mass Function}

In most astrophysical environments where the stellar mass function can
be constrained, it appears to be relatively normal. A similar initial
mass function (IMF) appears to hold from low density dwarf spheroidal
galaxies to dense star clusters, such as R136a. However, as the
locations and conditions of star formation become more extreme, 
might the IMF change?

One well studied case is that of M82, where some data suggested a
deficiency of low mass stars within the starburst, while later work
implied a relatively normal IMF. In the latest round of this saga,
Smith and Gallagher (2001) find the dense super star cluster M82-F
appears to be $\sim$7 times more luminous than predicted for a model of
its age with a Salpter IMF. Evidently this object currently lacks the usual
population of low mass stars, a situation that also seems to occur in
the Galactic center Arches star cluster.

Even if IMF peculiarities in starbursts were limited to only some super
star clusters, they could have important observational consequences.
For example, the clusters would then be over-luminous for their mass
and boost the intensity of a starburst for a fixed astration rate. This
would also imply short lifetimes for the super star clusters as bound
objects. As a result, much of the evidence of a starburst could
dissolve in only a few Gyr.

\subsection{Starbursts, Tully-Fisher Relation, \& Global Properties}

M82 provides a useful example of the relationship between starbursts
and global properties of galaxies. Figure 3 shows an observed
Tully-Fisher (TF) diagram, including the approximate location of M82.
We have made no correction for internal absorption and so the K-band
luminosity of M82 is a lower limit. As expected, M82 lies well above
the standard Tully-Fisher correlation, but as it is involved in a
short-lived starburst, galaxies in this state should be rare.

It is then interesting to ask how M82 will evolve on this diagram?
Note that if M82 were on the correlation before its starburst and its 
velocity width does not change duringthe starburst, 
then the postburst galaxy, that must 
be more luminous than the pre-burst system, will lie above the TF
relationship mean. This presents a problem; post-burst galaxies should
be neither obviously peculiar (and thus excluded from normal galaxy TF
samples) nor particularly rare. Yet there is little evidence for an
asymmetric upwards scatter in the TF diagram that would result from
postburst versions of M82.

How can this be? The problem mainly occurs if the velocity width for a
galaxy like M82 does not change during a starburst cycle. For example,
if starbursts increase {\it both} the luminosity and rotation
velocity, then galaxies would move up roughly along the mean TF
relationship after a starburst. Large deviations would occur only when
the stellar luminosity was high, near the peak of a burst, as in M82.
This process, shown schematically in Figure 3,
is similar to the requirements set by theoretical models of
galaxy formation (cf. Steinmetz \& Navarro 1999, van den Bosch 2000). 
We should be able to test this type of model 
in the rich populations of small disk galaxies in the Local Supercluster,
which include objects covering the full range of starburst
evolutionary cycles.

\begin{figure}
\epsfxsize=4.3in
\epsffile{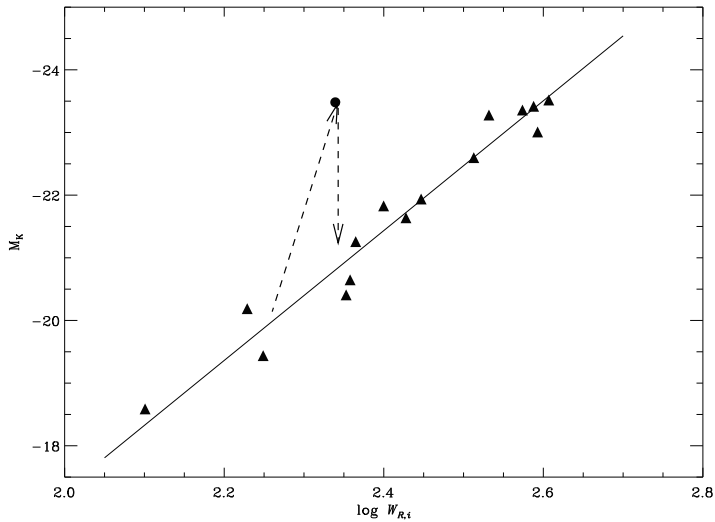}
\vspace{-4.1cm}
\caption{Infrared Tully-Fisher relationship showing the 
estimated location of M82 (filled circle) 
as compared with galaxies from the Verheijen (1997) Ursa Majoris 
"normal spiral" sample. The K magnitude for M82 is from Ichikawa et al. 
1995, and the dashed line shows an evolutionary path for M82 
that would allow it to remain on the TF relationship in its pre- and 
post-starburst phases.}
\end{figure}

\section{Conclusions}

Low-mass spiral 
galaxies with similar global properties, such as rotation speeds, 
sizes, or HI masses, can be found with star formation rates extending 
from $\la$0.1 to $\ga$1~M$_{\sun}$~yr$^{-1}$. The range in levels of star 
formation activity within these galaxies covers an even wider span. We 
observe examples of minimal star formation rates 
in the outer disks of some superthin Sd systems, where the gas disks probably  
are largely self-supported, and normal, feedback-dominated star formation 
in their central regions. 

At the other extreme, low-mass disks tend to be gas-rich and thus are prime 
targets for starbursts. When in this runaway mode of making new stars, 
large star forming `clumps' are common features, possibly resulting 
from the presence of huge, gravitationally bound gas complexes within 
their disks. The star-forming clumps are prime locations for the formation 
of super star clusters, which in turn could have unusual properties, such 
as top-heavy IMFs. 

Most starbursts in low-mass disks are associated with 
some form of collisional perturbation, which is likely to modify the mass
distribution within the starbursting system. Some form of dissipation to 
concentrate mass within the optical galaxy seems to be essential to preserve 
the Tully-Fisher relationship for postburst galaxies. Interactions may also 
play a role in converting low-mass disk galaxies from slowly evolving 
systems with superthin disks into more ragged and more active forms of 
late-type galaxies, such as the Magellanic irregulars.  

\bigskip
\acknowledgements
We thank NASA for support of much of this work through funding  
associated with various {\it Hubble Space Telescope} projects. 
JSG also acknowledges partial funding for this research by the National 
Science Foundation grant AST-9803018 to the University of Wisconsin and 
by the Vilas Trustees through the University of Wisconsin Graduate 
School. LDM thanks the National Radio Astronomy Observatory 
for her Jansky Postdoctoral Fellowship.

\bigskip

\leftline{\bf References}

\noindent
Barton, E. J., Geller, M. J., \& Kenyon, S. J. 2000, \apj, 530, 660

\noindent
Bergvall, N. \& R\"onnback, J. 1995, \mnras, 273, 603

\noindent 
Dalcanton, J. J. \& Bernstein, R. A. 2000, \aj, 120,203

\noindent
Elmegreen, B. G. \& Parravano, A. 1994, \apj, 435, L121

\noindent
Elmegreen, B. G. \& Efremov, Y. N. 1997, \apj, 480, 235

\noindent
Gallagher, J. S. \& Hudson, H. S. 1976, \apj, 209, 389

\noindent
Gallagher, J. S. et al. 2001, in preparation

\noindent
Giovanelli, R., Avera, E., \& Karaschentsev, I. D. 1997, \aj, 114, 122

\noindent
Gnedin, N., Norman, M. L., \& Ostriker, J. P. 2000, \apj, 540, 32

\noindent
Goad, J. W. \& Roberts, M. S. 1981, \apj, 250, 79

\noindent
de Grijs, R., O'Connell, R. W., \& Gallagher, J. S. 2001, \aj, 121, 768

\noindent
Homeier, N. L. \& Gallagher, J. S. 1999, \apj, 522, 199

\noindent
Howard, S., Keel, W. C., Byrd, G., \& Burkey, J. 1993, \apj, 417, 512

\noindent
Ichikawak, T., Yanagisawa, K., Itoh, N., Tarusawa, K., van Driel, W., 
\& Ueno, M. 1995, \aj, 109, 2038

\noindent
Matthews, L. D. 2000, \aj, 120, 1764

\noindent
Matthews, L. D. \& Gallagher, J. S. 1997, \aj, 114, 1899

\noindent
Matthews, L. D., Gallagher, J. S., \& van driel, W. 2000, \aj, 118, 2751

\noindent
Matthews, L. D. \& van Driel, W. 2000, \aaps, 143, 421

\noindent
Matthews, L. D. \& Gao, Y. 2001, \apj, 549, L191

\noindent
Mayer, L. et al. 2001, \apj, 547, L123

\noindent
Mihos, J. C. \& Hernquist, L. 1994, \apj, 425, L13

\noindent
Mulder, P. S., van Driel, W., \& Braine, J. 1995, \aap, 300, 687

\noindent
Noguchi, M. 1988, \aap, 203, 259

\noindent
Noguchi, M. 1999, \apj, 514, 77

\noindent
Schweizer, F. \& Seitzer, P. 1988, \apj, 328, 88

\noindent
Smith, L. J. \& Gallagher, J. S. 2001, \mnras, in press

\noindent
Smoker, J. V., Davies, R. D., Axon, D. J., \& Hummel, E. 2000, 
\aap, 361, 19

\noindent
Steinmetz, M. \& Navarro, J. F. 1999, \apj, 513, 555

\noindent
van den Bosch, F. C. 1999, \apj, 530, 177

\noindent
Vel\'azquez, H. \& White, S. D. M. 1999, \mnras, 304, 254
\noindent 

\noindent
Verheijen, M. A. W. 1997, The Ursae Major Cluster of Galaxies, Ph.D. thesis, 
Groningen University

\noindent
Wyse, R. F. G. 2000, in ``The Galactic Halo: from Globular Clusters to 
Field Stars",  in press.

\end{document}